# H is for Human and How (Not) To Evaluate Qualitative Research in HCI

Andy Crabtree, School of Computer Science, University of Nottingham, UK
ORCID ID 0000-0001-5553-6767
Email andy.crabtree@nottingham.ac.uk

**Abstract.** Concern has recently been expressed by HCI researchers as to the inappropriate treatment of qualitative studies through a positivistic mode of evaluation that places emphasis on measurement and metrics. This contrasts with the nature of qualitative research, which privileges interpretation and understanding over quantification. This paper explains the difference between positivism and interpretivism, the limits of quantification in human science, the distinctive contribution of qualitative research, and how quality assurance might be provided for in the absence of numbers via five basic criteria that reviewers may use to evaluate qualitative studies on their own terms.

## 1. Introduction

One of the reviewers of this paper suggested it should be called *A Beginners Guide to Reviewing Qualitative Research in HCI*. There is certainly something in this, insofar as it describes five criteria HCI reviewers might use to evaluate qualitative research, but there is more to it than that. Four of the five criteria obtain their purchase in being drawn from the interpretive tradition in human science, which is closely allied with qualitative research. In addition to furnishing criteria relevant to the evaluation of qualitative studies, this paper provides HCI researchers / reviewers with background knowledge needed to understand what kind of contribution qualitative studies make and why they should be evaluated in the ways recommended here. In short, it is not sufficient to provide HCI researchers with tools to evaluate qualitative research in methodologically congruent ways. It is also necessary for HCI researchers to understand some of the basics of human science and what is distinctive about it, so that they understand why these particular tools are appropriate. Much like any other tool, one needs to know something (but by no means everything) of how it works, what it does, what it doesn't do, etc.

Qualitative researchers may take issue with their studies being cast as a 'tool' and argue that their value does not necessarily lie in their utility. That may be true in human science, where emphasis is placed on understanding human phenomena, but qualitative studies have a broader role to play in HCI in helping researchers and practitioners identify and address substantive problems within the field (Oulasvirta & Hornbæk, 2016). Here they have to be useful. It is against this backdrop that evaluation of qualitative studies become problematic for reviewers. They often don't see how studies with small numbers of participants over short durations producing small datasets can be useful for HCI because they are so obviously limited. Rather than value qualitative studies for the unique insights they provide into the human world and examine their relevance to HCI, reviewers all too often resort to extraneous criteria, typically metrics and measurements: sample size, code counts, interrater reliability scores, etc. (Soden et al., 2024). The imposition of metrics and measurement on the evaluation of qualitative research is called 'positivistic' (ibid), an approach that is *incongruent* with qualitative research (Clarke et al., 2025).

The focus and contribution of this paper is on helping the HCI community remedy the problem of methodologically incongruent (positivistic) reviews of qualitative research. There is growing realisation within and without HCI that qualitative research is not being treated appropriately, on its own terms (e.g., Bardzell & Bardzell, 2016; Busetto et al. 2020; Soden et al. 2024; Clarke et al. 2025). This is not only bad for qualitative researchers, it's also bad for HCI insofar as it creates methodologically incoherent research practices that deprive HCI of crucial insights into the relationship between technology and society. The contribution turns on enabling HCI researchers / reviewers to appreciate the foundational difference between positivism and interpretivism, the limits of quantification in human science, the distinctive contribution of qualitative research, and how quality assurance might be provided for in the absence of metrics and measurement. This leads to consideration of five basic criteria, four of which as noted above obtain their purchase in being drawn from the interpretive tradition in human science, that reviewers may use instead of positivistic measures to evaluate qualitative studies on their own terms. These criteria orient HCI researchers / reviewers to methodological reflexivity, the apodictic nature of findings, the provision or elaboration of sensitising concepts, the analytic reach and potential utility of qualitative findings, alongside their relevance or usefulness to HCI.

## 2. Motivation and method

> "I have difficulty accepting the methodology. The results are based on two 1 hour discussions … with groups from 3 organisations and 3 roles … 5 people interviewed in total." Anonymous reviewer

The above quote was a comment on a qualitative study my colleagues and I conducted into practitioners' apprehension and treatment of risk in AI (Crabtree et al., 2024). The study was first submitted to a leading HCI journal and rejected on the grounds cited above. The rejection prompted the initial writing of this paper, including an account of my motivation (the inappropriate imposition of metrics to evaluate qualitative research), and was sent to the same journal. Despite receiving favourable peer reviews requiring minor modifications, "significant revisions" were required by a "sceptical" associate editor. I was told that the imposition of positivistic reasoning was not the problem. I was told that HCI suffers particularly from "rather shallow qualitative interviews-based work" that often lack "enough data", such as my own. I was told to "cut out the complaining" for while there is "a general challenge of poor reviewing in HCI" born of "ignorance", most HCI reviewers are "competent and act in good faith". Admittedly, they occasionally suffer "fatigue" and sometimes "misapply review criteria", but "that's life." The view that there is no problem with the way in which qualitative research is evaluated in HCI – negative outcomes are either a matter of shallow work or the occasionally ignorant or tired reviewer (bad luck, that's life) – stands in sharp contrast to the recognition that there is a *systemic* problem in HCI with the review of qualitative research. As Soden et al. (2024) put it,

> "We are concerned that increasingly all qualitative research, including that which draws on interpretive traditions of research design, is being evaluated from the perspective of positivism … This is not just a problem of the occasional Reviewer 2. In recent cycles of CHI and CSCW, numerous ACs have begun demanding such things, and these errors often slip past editors and others in the process whose role it is to catch them."

Soden et al. used social media to reach out to qualitative researchers in the HCI community and elaborate the point. Qualitative researchers reported that HCI reviewers often:

- Asserted that a study could not yield new insights because it only involves a small number of interviews or small amount of time doing observation.

- Asserted that a study is not representative and/or the findings aren't generalisable because the authors only conducted a small number interviews.

- Requested a table of demographic information about study participants, in a putative bid to ensure participants are representative of a broader population and the findings are therefore generalisable.

- Requested that participant IDs or pseudonyms are added to each quote to ensure they evenly distributed and avoid cherry-picking.

- Requested codebooks that include code counts and interrater reliability scores to ensure consistency and replicability of findings across teams as represented by high rates of agreement among coders.

As we can see, the 'positivist things' that Soden et al. speak of share a general concern with measurement, subjecting qualitative research to *evaluation by quantification:* how many, how much, how long, etc. Our sceptical associate editor reflects this in saying that interview-based work in HCI often lacks *enough* data, for example, just as our anonymous reviewer does in criticising the *duration* of data collection and *number* of participants.

The bad luck (that's life) argument contrasts with the suggestion that positivism is commonplace, widespread, and mundane in HCI. This is not to say that HCI researchers / reviewers think of themselves as positivists or could necessarily tell you what positivism means if asked, nor that many even recognise or actively identify with the term. But it is to say that many *practice* positivism on a mundane basis. That when we do such 'things' as assert that studies are not representative because they are based on a small *number* of interviews, or a short *duration* of study, or request *measures* to assess the representativeness of participants, or response *distributions*, or seek to assess the replicability of findings through interrater reliability *scores*, etc., we *apply* positivistic reasoning to qualitative research, whether we mean to or not, or do so in conscious recognition of what we do or not (unconscious bias). It is not simply that we often use quantification as a means of evaluating qualitative research that is of concern, it is also the (often unspoken) *reasons* for doing so, ingrained as Soden et al. observe through education and training. It might be argued, for example, that measurement is impersonal and thereby enables the objective determination of the representative nature of a study and with it the generalisability and replicability of findings. Quantification, in other words, is seen as an integral part of the scientific apparatus, key to the evaluation of research in HCI.

The positivist viewpoint contrasts sharply with the interpretivist stance on human science. Human science is a term that glosses a diverse array of disciplines: psychology, sociology, politics, geography, etc., disciplines that focus on the human world rather than the natural world (Human Science, 2025). The interpretive tradition within human science places emphasis on *understanding* the human world, in contrast to developing causal explanations as in the natural sciences, and on qualitative research as a fundamental method for developing our understanding. There are in practice many qualitative methods, though there appears to be little agreement as to what defines qualitative research across them apart from it involves an iterative process of investigation and understanding (Aspers & Corte, 2019). But an understanding of what? In laying the foundations of human science, emphasis was placed on "historical" understanding (Dilthey, 1977), which is to say that the human world and human conduct is to be understood from within the "socio-historical processes" (past and present) that shape everyday life and our lived experience. Socio-historical processes is rather dry way of talking about everyday life in all its richness. Seen from this perspective, metrics and measurement offer no guarantee of understanding. Indeed the suggestion is (as elaborated below) that they get in the way, imposing an inappropriate (incongruent) scientific ontology, epistemology and methodology on our understanding (Clarke et al., 2025).

What follows might best be seen as a 1-0-1 crash course that helps educate the HCI community and sensitise reviewers to what positivism and interpretivism are, the fundamental difference between the two, and what that difference means for evaluating qualitative research in HCI. My method or approach is socio-historical, as befits human science. It involves taking the reader back to the origins of human science and the sociological study of suicide in particular. Why? Because it is here that the *fundamental distinctions* between natural science and human science were drawn and the ontological, epistemological and methodological *foundations* laid. An appreciation of these foundations is important to understanding the unique contribution of qualitative research today and how to treat qualitative studies appropriately in HCI. This, then, is no idle history lesson. As Reeves (2015) observes,

> "If HCI is to be a rigorous inter-discipline (Blackwell, 2015) then it will require working more explicitly at the interface of disciplines. We will need *more* reviews of and reflections upon the landscape of different forms of reasoning in HCI and through this better ways of managing how potentially competing disciplinary perspectives meet together."

It is necessary to move away from the comfortable home ground of HCI and work explicitly at the *interface of human science* rather than computing, if we are to remedy the problematic treatment of qualitative studies.

**3. Positivism**

If you search up positivism on the internet you will find it characterised by words such as "experiments and tests", "observable, measurable facts", "objective truth", and you might be forgiven for thinking that positivism *is* science. It isn't. It's not scientific practice at all. It's actually a *philosophical* view on science . More than that, and perhaps surprisingly, it's a philosophical view on *human* science rooted in an appreciation of the early achievements of the natural sciences (astronomy, physics, chemistry, biology) and mathematics. Philosopher, mathematician and social theorist Auguste Comte suggested that the "primary objective" of human science was to "generalize our scientific conceptions, and to systematize the art of social life" (Comte, 1848). The systematizing and generalising were to be done by predicating human science on the "method of natural philosophy" (Comte, 1851) and thus on the development of observation-based theories that describe the laws that order objective reality and explain cause and effect. Comte developed an observation-based theory that described the "laws of (human and social) development", for example.

> "In my *System of Positive Philosophy* … I attempted, and in the opinion of the principal thinkers of our time successfully, to complete and at the same time co-ordinate Natural Philosophy, by establishing the general law of human development, social as well as intellectual … It lays down .. [that] all subjects whatsoever, pass necessarily through three successive stages: a Theological stage, in which free play is given to spontaneous fictions admitting of no proof; the Metaphysical stage, characterized by the prevalence of personified abstractions or entities; lastly, the Positive stage, based upon an exact view of the real facts of the case." (Comte, 1848)

The "exact view" Comte speaks of is very important to 'positive' philosophy and emphasizes the central role of *mathematics* to measure and quantify "the real facts of the case." Indeed mathematics provided the "true fundamental basis of this philosophy" (Comte, 1830). Positivism is, then, not natural science but a philosophy of *human* science that seeks to *mimic or ape* the natural sciences, observing social life through mathematical measurement and quantification so as to provide an exact view of the real facts of the case whose law-like operations are to be explained by the theoretical mind.

Comte never realised this positivistic ambition. It fell to others, notably Emile Durkheim, who developed a social explanation of what is widely perceived to be one of the most individual of human acts: suicide (Durkheim, 1897). Durkheim developed a 'positive' (cause and effect) theory of suicide by analysing exact observations of the real facts of the case: statistical records of suicide rates across Europe between 1841 and 1872. These allowed Durkheim to compare suicide rates within and between countries over time "in percentage form." Comparing eleven countries across three different time periods, he found that the "factual order" of suicide shows that "each society is predisposed to contribute a definite quota of voluntary deaths." Statistical analysis ruled out heredity, race, and the impact of the physical environment (weather) as well as normal and psychopathic psychological states. Durkheim thus concluded that "it must necessarily depend upon social causes." He went on to explain these social causes and their law-like operation, describing four types of suicide subject to the operation of two "coercive" social forces, "forces exterior" to the individual, having a "power controlling them." Thus, on the one the hand, egoistic suicide and altruistic suicide vary inversely with the degree of an individual's *integration* in society: too little integration results in excessive individualism, too much in insufficient individuation. Either way, the imbalance in social integration has "the same effects", the egoist "depends only on himself" and eventually finds existence wanting, the altruist "kills himself because it is his duty." On the other hand, anomic and fatalistic suicide vary inversely with the degree of moral *regulation* in society. Anomic suicide thus results from too little moral regulation, fatalistic suicide from excessive moral regulation "of persons with futures pitilessly blocked and passions violently choked by oppressive discipline (ibid.)." When the integrative and regulative forces are in "equilibrium", suicide rates are stable; when they aren't, suicide rates spike. Thus is cause and effect.

Few buy into Durkheim's explanation these days, fewer still Comte's theory of development. What endures is their positivistic take on *human* science and the idea that we can observe social life and determine the real facts of the case from the exact point of view of mathematics and the quantification of everyday life. Today, just about anyone can leverage a quantitative approach: design a questionnaire, get a statistically significant and representative number of people to complete it, then run the data through a spreadsheet, bespoke software or generative AI to analyse the variables and *voilà*, there are the social facts of the matter. Positivism is now commonplace and utterly taken for granted, even children learn it (e.g., BBC Bitesize, 2025), which of course is part of the problem qualitative researchers face insofar as it creates a certain ingrained, accepting and unquestioned way of thinking about the world. It was a radical idea at the time though, one that gave birth to social science.

**4. Interpretivism**
Equally radical was the alternate viewpoint that underpins qualitative research in the human sciences. In 1883, before Durkheim's *Suicide*, German philosopher Wilhem Dilthey published an *Introduction to the Human Sciences* (Dilthey, 1989). The preface excoriates Comte for trying to cast the study of the human world in the image of the natural sciences, as this can only "mutilate historical reality." While clearly anti-positivist, Dilthey was not anti-science, he was only against casting the human sciences in an image of natural science (and not necessarily a correct image). For Dilthey, natural science "overlaps" with the human world, it has practical connection with it (e.g., practical applications emerge from science and scientific knowledge informs human understanding, as we shall see with suicide), but human science nevertheless "forms an independent system" of knowledge, a system whose methods are different to the natural sciences due to an essential difference in the "content" of the human sciences. For Dilthey, "natural science analyses the causal nexus of nature", subjecting the natural world to "a process of calculation" and making it accountable to a "mechanistic framework". This framework is "incommensurable" with "socio-historical reality", which "provides the content" of the human sciences. Socio-historical reality consists and is the ongoing product of "systems of social life", "developed from the sphere of practical life" and the mundane business of *living together*.

> "These cultural systems endure while particular individuals themselves appear on the stage of life and disappear from it again … in each the content and richness of human nature … flows into such a system … shaped by the vital but transient activity of persons … [that] gives these systems an external permanence independent of the individuals themselves and the massive objectivity which characterises them." (ibid.)

Systems of social life – in contrast to the mechanistic workings of nature – and how they impact and are impacted by people are the focus and furnish the content of the human sciences. The "epistemological foundation" is different then: on the one the hand, the causal (mechanistic) workings of nature (which cannot be different), on the other, the practical workings of our ongoing social history and social systems (which can and are subject to reform). Thus, and as Dilthey put it, we cannot understand the human world by "seeking to force it into the Procrustean bed of knowledge of uniformities analogous to that in the natural sciences" (ibid.).

Dilthey drew the alternative out to positivism in further detail in 1894 when he explained his ideas for a descriptive and analytic psychology (Dilthey, 1977). Psychology, like sociology, sought to emulate the natural sciences (experimental psychology still does), reducing what Dilthey described as the "fullness of life" to "a system of causality." For Dilthey the fullness of life constitutes the "psychic nexus", which he says is the "basis of cognitional processes." The psychic nexus is not located in-the-head, however, but in the "different purposive systems, such as economic life, law, art and religion" that constitute our socio-historical reality. More prosaically we might say the psychic nexus is our society and culture with its distinctive forms of life, which "determines the nature of our understanding of ourselves and of others." Dilthey was therefore of the view that any study of the human world should "observe and collect everything which it can seize of the historical processes wherein such a nexus becomes constituted", including the past, present and ongoing "action of the physic nexus", as it is from within this nexus, and only from within it, that we humans *understand who and what we are* (ibid.).

Observation isn't to be done through mathematics and quantification, then, but by immersion in the socio-historical processes that articulate social systems, including the vital activities of persons, so that we might describe and analyse how they impact our understanding of human life: *no exact view then.* Importantly, analysis should "always preserve in it something of the living artistic process of understanding (Dilthey, 1977)."

> "In understanding we proceed from the coherent whole which is livingly given to us in order to make the particular intelligible to us. Precisely the fact that we live with the consciousness of the coherent whole, makes it possible for us to understand a particular sentence, gesture or action. All psychological thought preserves this fundamental feature, that the apprehension of the whole makes possible and determines the interpretation of particulars." (ibid.)

A simple example of what Dilthey is talking about here is provided by a signpost: on any particular occasion we interpret and understand a signpost because of the purposive system of transport including footpaths, bridal ways, and roads that give signposts their intelligibility. It is this whole cultural system or form of life that allows us to interpret and understand the particular instance (go *that* way) (Braver, 2019). It might otherwise be said that *understanding* relies on the use of *cultural knowledge* to interpret the particular things we encounter. For Dilthey this means that studies of the human world "must begin with culturally developed man" (Dilthey, 1977) and be rooted in the living art of interpretation that provides for his and of course her understanding of the human world. This renders positivism's objective stance problematic, as it means that what constitutes objective reality is always a matter of interpretation that relies on the perceivers' cultural knowledge (see Garfinkel et al., 1981 for an exemplary demonstration of the point).

### 4.1 Interpreting and understanding

Dilthey provides a very different orientation to, and way of thinking about, the study of the human world to Comte and Durkheim. Whereas the natural sciences observe and explain the casual workings of nature, the human sciences (and humanities) instead seek to *interpret and understand* social systems, how they work, and their impact on individuals (and increasingly our environment). This emphasis on interpretation and understanding was most notably incorporated into the foundations of the human sciences and qualitative research by Max Weber. Weber, like Durkheim, was a sociologist and saw its standing task as "the interpretative understanding of social behaviour in order to gain an explanation of its causes, its course, and its effects (Weber, 1962)." Now this may sound to the untrained ear like Weber is pretty much in agreement with Durkheim as to the business of sociology, but when Weber speaks of cause and effect he is not talking about the *same* thing: he is not talking about positivist explanations of the mechanistic laws governing human conduct, but about interpreting and understanding intelligible or meaningful human behaviour and its consequences.

> "Understanding may be of two kinds: first, direct empirical understanding of the meaning of a given act [i.e., from the point of view of parties to it] … [second], explanatory understanding … rationally based understanding of motivation … [which] can be accepted as true explanation of the actual course of behaviour."

> "Motivation as used here refers to … sufficient reason for his conduct (ibid.)."

When Weber speaks about cause it is *reasons* he refers to, and when he speaks of explaining effects he speaks of describing the outcome of "rationally based" – i.e., reason-able but not strictly logical – human conduct. Weber's seminal work *The Protestant Ethic and the Spirit of Capitalism* (Weber 1904, 1905) illustrates the point: the outcome of the Protestant ethic, which invested the pursuit of worldly gain with moral and spiritual meaning, was the social system we call 'capitalism'. No law-like operations here then. Instead, accountably rational conduct that furnishes the distinctive content of the interpretive human sciences. By accountably rational I mean the social systems that organise the human world *makes sense* to members, i.e., to the people who populate them, who inhabit them, whose activities and actions animate and give life to them, including those affected by them. This does not necessarily imply shared understanding or agreement, only that they everyday life is intelligible and describable to members and therefore accountable. Thus, and in contrast to mechanistic explanations, the human sciences trade in accountably rational descriptions of everyday life when they speak about and explain cause and effect in the human world.

Like Comte and Durkheim, it doesn't matter if one cares for Dilthey or Weber's explanations. What matters is their thinking laid the foundations for an alternative ontology and epistemology that replace quantitative explanations with qualitative interpretation and understanding of human conduct. While Weber was by no means the author of qualitative research, which has its professional roots in Chicago School sociology (Platt, 1985; Vidich & Lyman, 1994), his work came to shape the qualitative perspective following the publication by Alfred Schutz of his critically acclaimed work *The Phenomenology of the Social World* in the 1960s, which opened with "Max Weber's Basic Methodological Concepts" (Schutz, 1962). Weber's achievement isn't so much his seminal thesis, or his methodological writings, as the idea that professional study of the human world involves not one but two kinds of interpretation and understanding: direct empirical and explanatory. The human sciences are today rife with methods for obtaining direct empirical understandings (methods of observing and describing meaningful human conduct from the point of view of participants), and analytic frameworks for interpreting what is seen and providing accountably rational explanations are legion, their ongoing production the staple business of the human sciences (e.g., conversation analysis, interaction analysis, video analysis, thematic analysis, phenomenological analysis, etc.). There is, of course, disagreement as to whose methods and interpretations are right or better or more suitable, but that the right and proper job of qualitative research lies in interpreting the human world through direct empirical *and* explanatory understanding is nowhere disputed. Were it to be argued that direct empirical understandings are sufficient, that would mean one is representing the human world without understanding what is being represented or why (e.g., simply presenting a transcript or description of some action recorded on video). Some level of explanatory understanding is also necessary, an account which describes what is interesting about the direct empirical understandings observed and described, what they show, demonstrate, reveal, contribute (etc.) to this or that human science and understanding of this or that topic, otherwise we are just regurgitating the words and deeds of others. We may argue about what interpretation and explanation mean and how it gets done or whose analytic framework is better and why (Button et al., 2015), but not that qualitative research is fundamentally engaged with the business of interpreting and understanding the human world and human conduct past, present and, in HCI, future.

**5. Positivism and interpretivism**
Somewhat schizophrenically, positivism and interpretivism *co-exist* in the human sciences and present us with *fundamentally different* ontologies, epistemologies and methods of observation: on the one hand, positivism seeks to explain human conduct in causal terms and conducts observation through quantification and a battery of increasingly sophisticated mathematical techniques (Cicourel, 1976); and on the other hand, interpretivism seeks to explain human conduct by interpreting and understanding its accountably rational basis and conducts observation through an enormous variety of descriptive and analytic methods (Given, 2008). Some researchers, and this is not uncommon in HCI, combine the two and simply treat them as furnishing different kinds of *data* that can happily co-exist (e.g., Mozyrko, 2024). As Creswell (2009) points out,

> "Mixed methods research, employing the combination of quantitative and qualitative approaches, has gained popularity."

> "[However] qualitative procedures demonstrate a different approach to scholarly inquiry than methods of quantitative research. Qualitative inquiry employs different philosophical assumptions; strategies of inquiry; and methods of data collection, analysis, and interpretation."

Mixing quantitative and qualitative methods may be seen as useful and expedient, but one should exercise care and respect the fundamental differences between the two, otherwise the blurring of methods allows quantitative reasoning to seep into and colour qualitative research. It may seem obvious then, and for example, to ask of qualitative research how many participants, selected on what basis, observed over what period of time, etc., as if numerical reasoning were, quite naturally, an appropriate way to think about studies that seek to interpret and understand human conduct. Sociologist Jack Douglas draws out the dangers of doing so, however, in his qualitative study of suicide.

**5.1 Suicide revisited**
Douglas exposes the "fallacy of treating society *as if* it is somehow a separate level of existence, outside of the hearts and minds of live-and-breathing human beings" and assuming there is "only one general set of criteria for scientific validity or truth [which] is embodied in the classical works of the natural sciences (Douglas, 1966)." Douglas's study showed that, in contrast to the law-like operation of social forces à la Durkheim, suicide is the *outcome* of a "suicide argument process" in Western societies at least. Far from simply being able to read off suicide as a mode of death, "official categorizers [coroners] and others involved" – family members, friends, colleagues, first responders, pathologists, etc. – in the "interaction" and "communication" that constitutes the suicide argument process together "construct" the meaning of the deceased's actions and thereby determine cause of death. Suicide is, as Douglas describes it, "situationally problematic" as one cannot study it by "abstracting the communicators from concrete instances of suicide in which they are involved", e.g., by "questionnaires or laboratory experiments." Instead we must "look at the meanings imputed in concrete situations", as the suicide argument process unfolds. From this perspective,

> "any suicidal action is believed to mean something fundamental (or substantial) about the self of the individual committing it, or about the situation (especially the significant others) in which he committed the action, or about some combination of the self and the situation. Whether the specific meaning realized will be directed to the self or to the situation of the actor will depend on the imputations of causality made by the various interactors." (ibid.)

Imputations of causality turn upon "two general constructions" of meaning: insofar as suicide is considered a plausible option, which is not always the case (Tait & Carpenter, 2014), the individual is seen to be either "responsible" for their actions (e.g., they were depressed, insane, were terminally ill, sought to expiate their grief over the death of a loved one, etc.), or they were "driven" to it by the situation they found themselves in (e.g., loss of their job, family trouble, criminal proceedings, etc.). Importantly, the construction of such meanings explaining the cause of the deceased's suicide is "dependent upon much more evidence than simply his own statements", such as a suicide letter or prior threats, and turns on the "various interactors" interpretation of events: a wife, for example, who interprets her husband's actions as being a direct result of her filing for divorce or a toxicology report from a pathologist.

Douglas's study shows us that the social fact of suicide is not the result of impersonal social forces, but of live-and-breathing human beings engaged in an interpretive, interactionally produced, suicide argument process that is institutionalised in the West and generalises across Western societies. This *understanding* of suicide does not turn on number of observations, duration of study, number of participants or any other quantitative measure.

> "The problem with statistics on suicide … is not at all one of simply devising more accurate measurements. The fundamental problem is that determining and analysing the social meanings of suicide must be solved ***before*** one can attempt ***any*** quantitative analysis." (Douglas, 1966)

The failure of positivism is the failure to follow the empirical evidence "back to its actual source" (Douglas, 1970). When this is done one finds, as Douglas did, a very different world: a world inhabited by live-and-breathing human beings engaged in accountably rational human conduct and the production of social facts. Interpreting the human world does not turn on natural science. It may *overlap* to use Dilthey's phrase, e.g., through toxicology or pathology reports in cases of suicide, but as Douglas makes visible, "imputations of causality" turn on an interpretive process that is staffed by live and breathing human beings who seek to understand the motives and reasons the deceased might have to take their own life: motives and reasons that could make them responsible for such an action and/or drive them to it.

Positivism and interpretivism may co-exist in the human sciences – there are practitioners of both approaches today and as Cresswell (2009) points out, mixing them together has gained in popularity – but nevertheless *they do not overlap* as natural science overlaps with and is thus accountable to everyday life (e.g., as the pathologist's scientific analysis of death is accountable to determinations of suicide). It is easy to think that qualitative insights must be accountable to statistical patterns, but Douglas's work makes it visible that they are not. Indeed, that they are not even speaking about the *same* social phenomenon; suicide, for example. One treats it as a mathematical construct that reveals the law-like operations of impersonal and objective social forces on individuals, the other as an interpretive human process that organises the social determination of an individual's cause of death. They are not talking about the *same* thing at all. The only thing they share in common is a word and there is no overlap in their treatment of that word or the understandings they produce. One might argue that any given determination of suicide will map to Durkheim's causal types, but no one involved in the suicide argument process does and nothing in the suicide argument process turns on that typology. As Douglas notes, we can "study the phenomena of everyday life on their own terms" and "make use only of methods of observation and analysis that retain the integrity of the phenomena", or we can study them as phenomena "strained through" the classical methods of the natural sciences (Douglas, 1970). What we can't do is both, because it makes *no sense* to treat them as dealing with the *same* kind of phenomenon and to therefore assume quantitative reasoning, quite naturally, applies. Qualitative and quantitative studies are not *accountable to one another*, they do not overlap. They have *incommensurate* (or incongruent) ontological and epistemological foundations, methods and goals, and cannot, therefore, be judged by the *same* standards, criteria or rules (Winch, 1958).

## 6. Quantification and human science

There is intuitive appeal to the idea that quantitative reasoning is inherently scientific. Mathematics is the putative hallmark of science and scientific rigour, after all, so much so that it has become a *normative* feature of any activity that would wish to be considered science (Pérez-Escobar, 2023). With respect to the human world, however, Douglas was not convinced by what he called this "absolutist stance."

> "In general, the absolutist stance subsumed the everyday world under the methods of science and, in doing so, its users never realized that the everyday phenomena they observed were *scientified* phenomena."

Scientified sounds a bit strange to the contemporary ear. Today we might speak of scientism instead and the inappropriate application of scientific reasoning to the human world. As Gaver (2012) observes, "HCI is prone to scientism in its cultural assumptions" or the assumptions it holds as a community of practice. Reeves (2015) notes that among the community, HCI's status as an academic subject is "often in question, leading to desires for establishing a true scientific discipline." In the absence of foundational distinctions between positivism and interpretivism – no doubt a consequence of HCI being an inter-discipline (Blackwell, 2015) – mathematics and quantification would naturally appear to sit at the heart of this ambition. Thus, working under the normative assumption, it would make sense to subject qualitative studies to quantitative reasoning, because this is what science does, it's how science works. It isn't of course, it's how *scientism* works, driven in this case by the "disciplinary anxieties of HCI" (Reeves, 2015). Furthermore, it is a deeply ironic stance insofar as the quantification of human phenomena turns on *common sense reasoning*, not science at all!

### 6.1 Variable analysis

In 1958 Paul Lazarsfeld, sociologist and mathematician, introduced the human sciences to variable analysis (Lazarsfeld, 1958).

> "[Lazarsfeld] proposed that persons be treated as 'objects' displaying general properties … [e.g.,] age, gender, social class, status, intelligence, etc., as well as those pertaining to attitudes, opinions, beliefs and so on … [he] ingeniously deployed a metaphor from mathematics, the variable, to create 'devices' by which we can characterise the objects of empirical social investigation. A variable … can take one of a range of values which can be expressed numerically … as in the simplest of cases, where 0 signifies the absence of a property and 1 its presence. At a stroke, one of the classic objections to the quantification … of the human sciences, namely, that much of their phenomena are qualitative … was seemingly resolved in a practicable manner. Provided that an investigator could determine whether a qualitative attribute was present or absent, then it could be treated as a variable … Using the numerical values given to the variables, such as frequency counts, they can be mapped onto a coordinate property space … and made amenable to statistical partitioning in order to search for stable linkages in the data set." (Benson & Hughes, 1991)

This, of course, all sounds very scientific and exact, indeed Lazarfeld was of the view that variable analysis revealed "near causal relations" (Lazarsfeld, 1958). However, the quantification of persons and their general properties, attitudes and beliefs, etc., is rooted in common sense reasoning, *not* mathematics.

Variable analysis relies on the classification of social phenomena. Aaron Cicourel (1976) demonstrates how this works in reflecting on an example of the classification procedure provided by Lazarsfeld, which focuses on "why women buy a certain kind of cosmetic." Lazarsfeld tells us there are many reasons why women might do so: they value their appearance; they wish to make an impression on others; they appreciate the technical qualities of the cosmetics in question; they receive information from people they know or the media; they seek to avoid negative health effects; they are sensitive to cost; etc. Elaboration of the panoply of reasons implicated in buying a certain kind of cosmetic furnishes a "scheme of classification" that "matches the actual process involved in buying and using cosmetics" and thereby furnishes an array of "indicators" questions may be asked about. Cicourel subsequently observes:

> "The researcher relies upon his common sense knowledge … employ[ing] common sense concepts that reflect common knowledge known to both sociologist and the 'average' members of the community or society. By assuming from the outset that the social scientist and his subjects form a common culture which each understands in more or less the same way, the 'obvious' meanings of the operationalised questionnaire items on which the indicators are based, will incorporate properties … taken for granted as relevant to the research project." (ibid.)

The quantification of everyday life turns not on science but *common sense knowledge* and it does so, as Douglas makes visible in his explication of suicide, "for the simple reason that there is no other way to 'get at' the social meanings involved in social actions" (Douglas, 1970).

Does this negate the quantification of everyday life, an enterprise that has travelled far from sociology and become a staple feature of mundane discourse, the findings of which are regularly read in our newspapers or heard on our radios and TVs? Douglas didn't think so.

" … it is of great practical importance to us *as members of our society*. It is only some form of analysis of general patterns and structures that enables us to know what is going on across the far reaches of our social world so that we can take practical steps to try to control our social world in a way that … allows us to make rational choices that can affect all of our everyday lives (ibid.)."

That said, he was of the view that we should, *as analysts*, stop treating mathematical techniques that manipulate variables "*as if* they were scientific" given their demonstrable reliance on common sense knowledge and reasoning (ibid.). Quantitative studies of everyday life are not science, they are scientified, they tell us what age, and sex, and health, etc., look like after they have been strained through some coordinate property space, *not what they look like to live-and-breathing human beings*. Quantitative studies of the human world wear the mask of science, a respectable veneer provided by mathematics, but they do *not* trade in causes (mechanistic laws) as do the natural sciences and thus have *no* predictive power. Quantification in the human sciences deals with correlations, and it is well understood that correlations are not causes (Correlation & Cause, 204). Despite Lazarfeld's "near casual" claim, correlations are, at best, probably true. At worst, this is one of the primary reasons we routinely read about or hear false positive results in the mainstream media (Jarry, 2019). The veneer of science peels thin, so much so that the editor-in-chief of *The Lancet* suggested that "much of the scientific literature, perhaps half, may simply be untrue" (Horton, 2015). The quantitative will continue to co-exist with the qualitative in the human sciences, but there is no reason to see it as being *any more scientific*, or that increasingly sophisticated mathematical techniques will "overcome the epistemic gravity" of the situation (Thomes & Kneale, 2021). It is not science, it is not exact, it does not deal in cause and effect as the natural sciences do, but the *interpretation and understanding* of probabilistic relationships predicated on common sense reasoning.

**7. The unique contribution of interpretivism**
The human sciences and the natural sciences are different, by now that much should be clear: one deals with causal explanations of nature, the other with accountably rational explanations of the human world and human conduct. The former turns in important ways on exact mathematical measurement. The latter, even where mathematics are concerned, turns on interpretation and understanding. The human world is a pre-interpreted world, a world *already* replete with social meanings. Qualitative and quantitative research thus operate "from inside" society and are rooted in a common language and common body of understanding shared with its members (Cicourel, 1976). Quantitative research uses this common language and common body of understanding to identify and develop variables for mathematical analysis and to interpret probabilistic relationships between them. Qualitative research uses it to interpret and understand the socio-historical processes at work in society and its many and varied social systems and cultural forms of life. These processes may literally be historical or, as Douglas has shown us, they may be staffed as the suicide argument process is staffed by live-and-breathing human beings.

Across the human sciences, from descriptive and analytic psychology to sociology, anthropology, cultural studies, human geography, political science, economics, etc., qualitative researchers are seizing what they can from the socio-historical processes that make up society through a diverse portfolio of qualitative methods. The result, the output and contribution of qualitative research, is the production of *analytic* insights that *sensitise* us to important social and cultural features of everyday life and how live-and-breathing human beings experience them.

## 7.1 Sensitising concepts

The idea of sensitising concepts was developed by sociologist Herbert Blumer, who was of the view that concepts are the means by which we establish a connection with the empirical world.

> " … the concepts of our discipline are fundamentally sensitising instruments. Hence, I call them 'sensitising concepts' and put them in contrast with definitive concepts … A definitive concept refers precisely to what is common to a class of objects, by the aid of a clear definition in terms of attributes or fixed bench marks … [which] serve as a means of clearly identifying the individual instance of the class and the make-up of that instance that is covered by the concept. A sensitising concept lacks such specification of attributes or bench marks and consequently it does not enable the user to move directly to the instance and its relevant content. Instead, it gives the user a general sense of reference and guidance in approaching empirical instances." (Blumer, 1969a)

Sensitising concepts are not 'umbrella constructs' that loosely encompass and account for diverse phenomena (Tractinsky, 2018). As Blumer makes clear, they orient the user to general features of the human world and provide guidance for their elaboration in specific empirical instances. The concept of 'articulation work' which is foundational to CSCW, for example, orients and guides the user to attend the ways in which people "fit lines of action" and "activities to one another" in order to "mesh" them together and accomplish cooperative work (Schmidt & Bannon, 1992). For examples of empirical instances of articulation work see Heath & Luff (1991), Hughes et al. (1992), Button & Sharrock (1997).

It is not uncommon for qualitative research to be criticised for failing to provide hard and fast definitions of its concepts, but as Blumer makes clear, qualitative research *does not* trade in definitions. Rather, in providing sensitising concepts qualitative studies orient us to important social or cultural phenomena, phenomena that "shape up in a different way in each empirical instance" and require us to "work with and through the distinctive or unique nature of the empirical instance" (Blumer, 1969a), just as the empirical instances above do in elaborating articulation work in London Underground control rooms, air traffic control, and the print industry. In place of definitions, sensitising concepts are articulated through descriptive "exposition which yields a meaningful picture, abetted by apt illustrations." Their description and empirical elaboration furnishes *corrigible sketches* of social phenomena, not exact measurements or probabilistic relationships. *More* can always be said and they may as such be extended, developed or "improved" through continued exposition enabled through qualitative research and "patient, careful and imaginative life study" (not mathematics).

We are not suggesting that all qualitative research is rooted in Blumer's thinking any more than it is in Douglas's or Weber's or Dilthey's, only that in one way or another it taps into the ongoing flow of socio-historical processes and seizes what it can, as its methods provide, to furnish and / or elaborate sensitising concepts that orient us to important social or cultural phenomena and members lived experiences of them.

> "The history of qualitative research reveals that the modern social science disciplines have taken as their mission the analysis and understanding of the patterned conduct and social processes of society" (Denzin & Lincoln, 1994)

Similarly, we do not suggest that all qualitative research presents its outcomes as furnishing or elaborating sensitising concepts, only that the notion of sensitising concepts provides HCI with a useful heuristic to understand the *unique contribution* of qualitative research. Importantly, sensitising concepts are possessed of three key interrelated characteristics. 1) They introduce "a new orientation or point of view"; 2) they are an instrumental tool and function as a "fashioner of perception"; and 3) as such, make possible "the anticipation of new experience" and further investigation by others (Blumer, 1969b). Reflecting on the importance of sensitising concepts to the human sciences, Blumer was of the view that "the introduction of precision devices" such as the variable was "working along the wrong direction." The challenges confronting the human sciences do not stem "from the inadequacy of our techniques" but rather "from the inadequacy of our point of view." Sensitising concepts are crucial to developing an adequate viewpoint on the human world and human conduct. They sensitise us to it, orient us, motivate our investigations, and ultimately help us interpret and understand the world around us. It is important to appreciate that sensitising concepts may initially be quite "crude", that they are corrigible and "become refined" over time in becoming social concepts investigated by the research community (ibid.). The notion of articulation work, for example, oriented qualitative studies to its elaboration in lived human experience and the conduct of specific human jobs carried out by specific people in specific locations (no quantitative selection criteria here), which in turn furnished further sensitising concepts, such as distributed coordination, plans and procedures, and awareness of work (Hughes et al., 1997). The imposition of quantitative reasoning on qualitative studies is not only inappropriate, then, it may also quash innovation through improper measurement (how many, how much, how long, etc.) just as a thirst for precise definition may quash new orientations and points of view on society and technology.

## 8. Quality assurance

Positivism and interpretivism present two *fundamentally irreconcilable or incongruent* ontologies, epistemologies and methodologies. Ontologically, the world is objectively given for positivism, whereas for interpretivism what is given is always a matter of interpretation and understanding and turns on our cultural knowledge. Epistemologically, the world is known through exact measurement for positivism (e.g., variable analysis), whereas for interpretivism it is known through descriptive analytic practices. Methodologically, the world is understood through causal (or near causal) explanation for positivism, whereas for interpretivism it is understood through analytic frameworks that explain the accountably rational (intelligible) character of the human world and human conduct. In the human sciences, positivism trades in variables and probabilistic relationships, whereas interpretivism trades in culture as experienced by live-and-breathing people and concepts that sensitise us to that. The two are not at all the same. They deal with different phenomena in irreconcilably different (incongruent) ways. They are not accountable to one another. They are incommensurate. One cannot *sensibly* apply quantitative reasoning to qualitative research without riding roughshod over its own substantive base. So how are HCI reviewers – and I speak particularly but by no means exclusively to reviewers (including associate editors) of large conferences such as CHI and CSCW and leading journals such as HCI and ToCHI where there is a notable problem with the treatment of qualitative research – to know if the qualitative studies they are presented with are OK? How can reviewers, especially those unfamiliar with qualitative research, be assured of their quality if the numbers don't stack up and they aren't allowed to try and stack them up?

### 8.1 Reflexivity

Reflexivity is a definitive component of third wave HCI (Bødker, 2006) and the turn to human-computer interaction as phenomenologically situated (Harrison et al., 2007). It is rooted in the critique of objective representational practices (e.g., Gaver, 2012) but is not reducible to a single concept. As qualitative researchers have elsewhere observed (Olmos-Vega, 2023), there are many different types of reflexivity. Lynch (2000), for example, identifies over twenty different reflexivities in the human sciences. The *methodological* reflexivities are relevant here and include philosophical self-reflection, methodological self-consciousness, and methodological self-criticism. Philosophical self-reflection involves an inward-looking, self-critical examination of one's own beliefs, assumptions and bias. Methodological self-consciousness would have the qualitative researcher apply philosophical self-reflection to their *relationship* with the people they study (the so-called "other") and *develop* conscious awareness of how they position themselves towards and treat the other as subjects or objects of study, including the problems encountered in doing so (Hurst, 2023). Methodological self-criticism takes the next philosophical step and would have the qualitative researcher adopt a critical stance towards their relationship with those they study and make it *explicit* how they have arrived at understanding of them warts and all. There is nothing radical about methodological reflexivity, as Lynch points out "standard conceptions of science emphasize systematic self-criticism" (Lynch, 2000). What is different is the way in which methodological reflexivity plays out in qualitative research. As one might expect, methodological reflexivity involves the provision of a methodological account, which distinctively in qualitative research explains the 'positionality' or relationship of the researcher to the participants (Leigh & Brown, 2021) and the reflexive techniques employed (Markham, 2017). More radically, it may also lead to writing strategies that seek to make the *production* of social and cultural knowledge *within* the researcher-researched relationship perspicuous (Dourish, 2014). In HCI, for example, defamiliarisation (Bell et al., 2005) and confessional and impressionistic (Rode, 2011) writing strategies have been employed to develop reflexive texts that make it accountable how qualitative researchers *came to know* the social and cultural forms of life they studied *in* their relationship with those they studied.

Methodologically reflexive texts account for the *relational production* of social and cultural knowledge, i.e., how the qualitative researcher arrived at an understanding of "the other" in and through their *social relationships* (interactions if you prefer) with them. The idea that knowledge is essentially a co-production, the outcome of a qualitative researcher's interactions with the participants in a study, runs counter to positive conceptions of objectivity in which knowledge is the outcome of an impersonal, exact, generalisable and reproducible scientific method. Nonetheless, and it is a point that HCI reviewers should take seriously, the validity of qualitative studies does not turn on measurement, on re*counting* how many, how much, how often, how long, etc. Measurement does nothing to assure us of the validity of a qualitative study or its findings, as it has *no relation* to the social relationships in and through which the knowledge was and is produced. Quality assurance resides instead in the *practical adequacy* of the reflexive account, i.e., the description of the relational production of knowledge, which is to say a reviewer should be able to read an account of a qualitative researcher's methodology, understand how they interacted with the participants, and how those interactions produced the knowledge (findings, insights) provided.

**8.2 Apodicity**

Apodictic means that the insights produced by a qualitative study are self-evident. Qualitative researcher's document their observations and use data extracts to support their interpretations, "abetted by apt illustrations" as Blumer put it (Blumer, 1969a). Thus texts of all kinds (transcripts, diary entries, social media, etc.) sit alongside and are often woven together with visual media (drawings, photographs, videos, etc.) to elaborate the qualitative researchers understanding of social and cultural phenomena and to substantiate the results of their studies. When done well, the use of data not only provides a solid evidential base for analytic claims, it furnishes "vivid exhibitions" (Crabtree et al., 2012) that render findings *apodictic*. By this we do not simply mean they are, as per dictionary definition, clearly established or beyond dispute. Drawing on Edmund Husserl's notion of apodicity, we mean that they are so because data analysis, which involves the interpretive use of data to understand and represent the other's world, renders them *self-evident*. Husserl suggested that any rigorous understanding of the human world must be "grounded", not in Glaser & Strauss's sense of the word which emphasizes "the discovery of theory from data" (Glaser & Strauss, 1967), but in the sense that our understanding stands "upon a foundation of immediate knowledge whose self-evidence excludes all conceivable doubt" (Husserl, 1970). The axiomatic character of mathematics renders it apodictic, for example, but as Dilthey reminds us "there are not only numerical relationships which have this apodictic character" (Dilthey, 1989). We may not personally have witnessed or experienced Douglas's suicide argument process, for example, but most competent adults in Western cultures would accept Douglas's account as being apodictally true.

Nonetheless, reviewer confidence in qualitative findings is easily shaken, as the following extract taken from a recent HCI review illustrates:

> "It is also clear that many of the points raised are ... self-evidently important … [however] the validity of the findings presented is compromised by the narrow methodological approach, relying solely on a limited set of interviews … This raises questions about the reliability of the insights and whether they can be confidently applied by researchers and practitioners in HCI … incorporating a mixed-methods approach or a larger, more varied participant pool could enhance validity." (ibid.)

It is this kind of practice that needs to stop. Reviewers should feel empowered and have confidence in findings that are "self-evidently important" and thus apodictally true. They need not, and should not, appeal to extraneous and essentially *arbitrary* quantitative criteria. How many interviews is adequate, for example, with how many participants? Will the same numbers apply next time? On what basis would they be different? As Busetto et al. (2020) note when assessing the treatment of qualitative research in the health sciences, "being qualitative research instead of quantitative research should not be used as an assessment criterion", nor should qualitative research "be required to be combined with quantitative research per se." As Denzin and Lincoln (1994) explain, mixed-methods approaches are direct descendants of "classical experimentalism", positivist through and through. They take qualitative research out of its natural home within the interpretive framework and make it "auxiliary" to quantification. In doing so findings that are apodictic and self-evident are dismissed in being rendered accountable to the extraneous logic of quantitative science.

**8.3 Objectivity, generalisation and reproducibility**
It might be expected, at least by the HCI reviewer unfamiliar with qualitative research, that the findings it offers are objective, generalisable, and reproducible as these are putative hallmarks of science. Yes, but no. Let's start with reproducibility. As Reeves points out,

> " .. the replication of results as a necessity in principle for HCI has been described as 'a cornerstone of scientific progress' .. Yet when we turn to studies of the natural sciences, from which the principle is ostensibly extracted, we find that firstly most scientific results do not get replicated and secondly that where it is used, replication is a *specifically motivated*, pragmatic action for particular contested, relevant cases." (Reeves, 2015)

The natural sciences rarely replicate or reproduce their results, it would be extremely time-consuming and costly to do so, and it assumes others share the same technical apparatus to do so, which is often not the case. Rather, and as Reeves makes clear, replication has to be warranted, "specifically motivated." In other words, there has to be good reason to do so, such as significant doubt in the results, where the results are themselves significant and consequential. Reproducibility is *not* a standard requirement of scientific practice, only an *occasional* one, and it is not a requirement of qualitative research at all given that knowledge is a *relational* product and thus very much *dependent* on the individual researcher(s) who carried out the study. Apodicity not reproducibility is what counts, which is to say that confidence in the results of quantitative research should not be undermined by questions of reproducibility or replication. Such questions are not definitive of scientific inquiry and do nothing to assure the quality of a qualitative study. What then of generalisation?

*8.3.1 Generalisation*
Sociologist Harvey Sacks wondered how it was that anthropologists' procedures that ask but "one or two people more or less extended questions" produce results that often turn out to be generalisable (Sacks, 1992)? His answer was that a) the members of a society are embedded in a culture, b) culture is "orderly", it consist in particular arrangements of mundane activity and ways of doing things, and c) the slice of culture members experience in their formative years provides them with an "adequate sampling set up" that enables participation in social life. Sacks observes that any member encounters from their infancy only a very small portion of their culture, and a *random* portion in many respects given "the parents he happens to have, the experiences he happens to have, the vocabulary that happens to be thrown at him in whatever sentences he happens to get", but this sample nevertheless sets them up much like everybody else as "workable things in a society" able to deal with pretty much anybody else and thus engage in the mundane business of everyday life. There are exceptions, of course, but for the most part the sampling set up suffices. The anthropologist may find that they get enormous generalisability from but one or two people then because "for a member encountering a very limited environment, he has to be able to do that, and things are so arranged as to permit him to."

> "For example, we might take it as a tremendous puzzle, how it is that Whorf, studying the language of the Navajo, talking to one in New York, could build a more or less decent grammar of it .. then say 'Isn't it fantastic. What genius' .. But then we could say, well, how many people does any given Navajo encounter when he learns to do Navajo?" (ibid.)

It's not the case that qualitative research doesn't generalise then, but that it does so on a *different basis* to positivism. As Soden et al. (2024) note. many of the requests qualitative researchers get from reviewers are for metrics of one kind or another – sample size, demographic information, code counts, interrater reliability scores, etc. – in an effort to ensure study participants are "representative of a broader population" and that findings are therefore generalisable. However, and as Sacks makes clear, generalisation in the human sciences doesn't work on the bases of metrics and representative samples (other than in quantitative forms of analysis). It works on the basis of a society's *cultural arrangements*, including its language. Generalisation has nothing to do with individuals then, other than that their sampling set up enables them to apprehend, respond and elaborate when asked (or observed) the cultural arrangements that organise everyday life in their society. Some of those arrangements may operate at scale and even be ubiquitous (e.g., eating, commuting, shopping), whereas others may be much more local and contained (e.g., deviant sub-culture or the culture of work in a specific organisation). There are limits to generalisation then, even in the face of globalisation, but this is not to say that findings that do not generalise between cultures or within a culture are, therefore, of little value. The notion of 'articulation work' derives from qualitative research in a hospital (Strauss et al. 1989), for example, but had enormous *analytic reach* shaping the field of Computer Supported Cooperative Work (Schmidt & Bannon, 1992). Similarly, the notion 'situated action' had an enormous impact on HCI, and is derived from qualitive studies of human interaction with prototype photocopiers at Xerox (Suchman 1987). Generalisation is not a measure of any importance in qualitative research, it does not require it. What matters is the *analytic insight* provided into the human world and human conduct and, where the evaluation of qualitative studies in HCI is concerned, its *potential utility and reach*.

*8.3.2 Objectivity*
While we're dispensing with shibboleths, let's dispose of objectivity too or at least that version of it that holds that there exists an objective world that can be known independently of the knower through exact and impersonal measurement. As Weber noted, the human sciences operate from within a pre-interpreted world (Weber, 1962), the socio-historical world that is home to Dilthey's "culturally developed man" (Dilthey, 1977). They are, as such, "second order" disciplines (Schutz, 1962) that depend on a social world known in common by members to identify study topics, and rely on the use of mundane cultural understandings and concepts (such as suicide, crime, delinquency, etc.) as resources for analysis (Dennis, 2019). That everyday life furnishes topics of inquiry and common sense knowledge basic analytic resources, and not only for variable analysis but every analysis of the human world, means that there is *no possibility of objectivity* as understood by positivism in the human sciences (Bittner, 1973). We have no option but to rely on our subjective, or more accurately *intersubjective* (known in common) knowledge of social reality. However, this should not be seen as problem for the analyst, whether acting in their capacity as a member, a human science researcher, or reviewer of a qualitative paper in HCI. As Sacks reminds us, "on the matter of subjectivity, of using a member's knowledge and access" to everyday life, "the issue has, by and large, never been one of not using it at all" in the human sciences, the issue has been what to do with it (Sacks, 1992). For Sacks, members' knowledge was and is the rightful topic of the human sciences, but there is *no time out* from it being a resource too and no possibility of evasion (Garfinkel & Sacks, 1970). We might, then, usefully employ a mundane version of objectivity, a common sense version, which makes no assumptions about social reality but instead would have us set our personal bias aside, including our assumptions about science and how it works, and treat the work in front of us fairly, evaluating the evidence impartially, without being unduly influenced by our personal feelings. At the same time, we might as Soden et al. (2024) suggest, also acknowledge the limits of our knowledge and refrain from trying to assess approaches that are *outside* our expertise, at least without appropriate guidance. It's OK to say I don't know. It's not OK to subject a qualitative study to quantitative criteria of evaluation because you don't know.

**8.4 How to assess qualitative research**
It is time the use of erroneous scientific reasoning to assess qualitative research in HCI stopped. It is a serious form of *abuse*, it does *harm* both to individuals whose work is subject to it and to the areas of qualitative research they work in, suppressing innovation and hampering career development, and it is detrimental to HCI too. As Soden et al. (2024) put it,

> "If this lack of understanding is left unaddressed, we may be robbing HCI of the very forms of research and insights that are arguably best suited to help address many of our most pressing concerns around the relationship between technology and society."

It is twenty years since Denzin and Lincoln (1994) urged qualitative researchers to "resist conservative attempts to discredit qualitative inquiry by placing it back inside the box of positivism." The resistance is pretty much over in the human sciences, interpretivism has superseded positivism. It was found to be deeply conservative, its theories and methods incapable of apprehending social reality and responding to social change (Sharrock et al., 2003). It is no longer a dominant force in the human sciences, though retains a "ghostly presence" in post-positivist thinking (Hammersley, 2019). HCI, as an inter-discipline (Blackwell, 2015), inevitably lags behind. Insofar as HCI wishes to employ qualitative approaches, it is time it caught up and recognised the impropriety of subjecting them to positivist reasoning. Asking how many, how much, how long, etc., is as Denzin and Lincoln (1994) put it "an attempt to legislate one version of truth over another." I would urge the HCI community to accept, as it has long been accepted in the human sciences, that *multiple* versions of truth are possible and may co-exist without irony (we are dealing with people, not atoms, etc.), and to treat qualitative research on its own terms with the respect it deserves. So instead of asking how many, how much, how often, how long or for any other *metric or measurement* and thereby attempting make qualitative research accountable to positivistic methods and reasoning, attend instead to some of the *basics* of qualitative research.

Reviewers of qualitative research should recognise that the human sciences are second order disciplines and that knowledge is a relational product that very much depends on the individuals involved. They should therefore replace any concept of scientific objectivity with mundane objectivity, set aside personal bias and feelings, assess a study impartially, and ask and answer questions addressing these *five basic criteria*:

- **Is the methodological account provided adequately reflexive?** Knowledge is not objective in qualitative research, it is relational, the product or outcome of the researchers' interactions with participants. A methodological account should be provided that enables the reader to understand how the researchers interacted with participants so as to produce the knowledge (findings, insights) described in a study.

  *Applying the criteria:* Ask yourself is a methodological description provided? Does the methodological account describe: i) The researchers' relationship to the participants? ii) How the researcher treated participants in the course of conducting the study, i.e., the practical ways in which they engaged participants in the research (including salient problems encountered and how they were resolved)? iii) What data was collected and how it was gathered? iv) The analytic approaches employed to interpret the data and understand the participants?

- **Are findings apodictic (self-evident)?** Data (transcripts, videos, diary entries, social media, etc) is not only an object of analysis in qualitative research, it is a representational tool, drawn upon in an interpretive act to elaborate and substantiate findings. When done well, the interpretation of data provides 'vivid exhibitions' that render the analytic insights offered self-evident and beyond conceivable doubt.

  *Applying the criteria:* Ask yourself is data provided? Does it illustrate key findings? Are findings immediately available, i.e., can you see how they were arrived at from the interpretation of the data presented? Are the findings beyond conceivable doubt, i.e., does the interpretation of the data support and convince you of them? Reviewers should not try to conceive of ways to doubt the results of a study, however, or imagine how they could be different. They should focus on the apodicity of the results presented and whether or not they are clearly supported by the data or there are unexplained gaps between data and findings.

- **Does the study elaborate, extend or refine an existing sensitising concept or furnish a new one**? The unique contribution of qualitative studies is the provision, elaboration, extension and / or refinement of sensitising concepts, analytic descriptions that orient us to important features of the human world and human conduct including the role and impact of technology within it.

  *Applying the criteria:* Ask yourself is the study based on an existing sensitising concept or does it introduce a new one? If it is based on an existing one, is it clearly presented and can you see how the study elaborates, extends or refines it? If it introduces a new sensitising concept, does the study provide a new orientation or point of view on the human world and / or human conduct? Reviewers should not reject qualitative studies because the sensitising concepts they furnish are incomplete. Rather, they should ask if they have instrumental value and could therefore function as a 'fashioner of perception' that might attract the broader interest and attention of the HCI community. Reviewers should not reject qualitative studies because they do not provide definitions of sensitising concepts either, but rather ask if their description is sufficient to orient others to their investigation and elaboration?

- **Do the results have analytic reach and actual or potential utility?** The potential instrumental value of a qualitative study means that qualitative studies may provide analytic insights that are useful to HCI. Reviewers should seek to determine if they are or if they have the potential to be usefully employed in understanding some aspect of human-computer interaction in contemporary life.

    *Applying the criteria:* Reviewers should substitute concern with quantitative means of generalisation and the replication of findings and ask instead whether or not a qualitative study furnishes analytic insights that are useful to HCI or could be useful to HCI? Regardless of sample size, duration of study, amount of data gathered, etc., does the study shed light on a novel aspect of human-computer interaction or new light on an established topic? Could the analytic insight provided extend beyond the specific people studied and have greater reach and utility in HCI? Does it or could it potentially help the HCI community understand social and cultural aspects of computing more generally?

- **Is the study of relevance to HCI?** The question of relevance is fundamental to reviewing in any field. In HCI, the question is closely coupled to the actual or potential utility of qualitative study to the field. It may be possible that a reviewer can see the relevance or potential relevance of a study to HCI, but that the case has not actually been made in the study report (paper) or not been made well.

    *Applying the criteria:* Reviewers should ask if the relevance of a qualitative study to HCI has been stated and is adequate? Is a clear case provided in presenting, discussing and concluding the study to establish and underscore its relevance to HCI and how it in addresses a substantive problem within the field? Is related work in HCI explained and referenced to support the case and elaborate the substantive problem? In the case of absolute novelty, has the potential relevance to HCI been established?

These questions provide a basic evaluative framework for assessing qualitative studies in HCI. They are built on the foundations of human science, rather than an incongruent philosophy that would model human science on natural science. These foundations, as we have seen, respecify the ontological view of what human science is about, transforming it from an enterprise that seeks to *explain* the human world in mechanistic (cause and effect) terms, to one that seeks to *understand* the accountably rational character of everyday life. They respecify the epistemology of human science, shifting the production of knowledge from exact and impersonal measurement, to immersion in socio-historical processes and vital (as in lively) activities that articulate social systems. They respecify the methodology of human science and how we come to know the human world from a process of specifying and interpreting the relationships between mathematical variables, to one dependent on the researcher's relationship with participants and the development of sensitising concepts. They respecify general scientific criteria, dispensing with replication as irrelevant to the validity of either natural or human science, reframing generalisation as analytic insight, and objectivity as impartiality rather than a position that contrasts in principle with the very personal and subjective nature of qualitative research. Instead, HCI reviewers are urged to attend to the adequacy of methodologically reflexive accounts, the apodicity of findings, their ability to sensitise the HCI community to important social and cultural phenomena implicating technology, the analytic reach and utility of results to HCI, and whether or not qualitative studies introduce or help HCI address substantive problems in the field. The foundations of human science provide HCI research with criteria sufficient to evaluate qualitative research on its own terms. Reviewers need not turn to extraneous measurement.

Are the criteria complete? Could more be said? Could they be further elaborated, extended, refined? Could other criteria be considered? If the reader has understood anything about qualitative research in what has been said so far, then the answer is 'you bet'. They are not complete. They are not definitive. They never will be. Human science and ~~are~~ ways of understanding the human word are constantly evolving. The criteria provided here are a place to start remedying the current problematic situation in HCI, not an end. Other frameworks are available. Small and McCrory Calarco (2022) for example, provide a quality framework for assessing qualitative studies that turns on various criteria including self-awareness, cognitive empathy, and the palpability of participant experience.

And alternative or contrasting solutions might be also appealed to, which examine whether or not a qualitative study fits a particular genre (Sarker et al., 2018), for example, or which interrogate the "position" of humanistic writings "as contribution" (Bardzell & Bardzell, 2016).

> " … a position is not merely a proposition; instead, it holistically comprises an expert subjective voice, a theoretical-methodological stance, its own situatedness within a domain, a pragmatic purpose, and one or more propositions." (ibid.)

One could treat this paper as furnishing an example of position-as-contribution. The outcome of the position that it is inappropriate to subject qualitative research to quantitative forms of evaluation, is the proposition that the evaluation of qualitative studies would be better served through the application of the five basic criteria outlined above. These criteria furnish reviewers unfamiliar with qualitative research the tools they need to assess qualitative studies with the confidence that they are treating them appropriately. What validates them? Two hundred years of social thought. That's why we went back to basics, to the origins of human science. So the reviewer understands what is distinctive and unique about it and why the five criteria presented above make sense and are sufficient for the practical purpose of reviewing qualitative research in HCI. They are, nonetheless and to reiterate, merely a place to start. As Reeves (2015) notes, "we will need *more* reviews of and reflections upon the landscape of different forms of reasoning in HCI" and we should expect these reviews and reflections to further elaborate appropriate criteria for assessing qualitative research on its own terms.

## 9. Conclusion

A great deal may be had from qualitative studies that last days, hours, and even minutes. They don't require large sample sizes, long durations of study, large quantities of data, or need to be mixed with or made accountable to quantitative methods and philosophies of natural science. The view that qualitative approaches simply provide HCI with methods of data collection is also deeply misplaced. The reduction of qualitative research to mere method dispenses with the very thing that makes qualitative research what it is. Gaver et al. (2004) made a similar observation in discussing HCI's treatment of cultural probes, noting the "strong tendency to rationalise probes" and how this "dilutes" the approach and ultimately "misses the point." The point of qualitative research is not merely to provide methods with which to gather and analyse data, but to *understand* the human world and human conduct. The rationalising tendency of HCI dispenses with the standing task of the human sciences and makes qualitative research accountable to a deeply ingrained positivistic mode of reasoning that seeks assurance in measurement and metrics. It is cultural appropriation gone badly wrong. As Dourish pointed out with respect to ethnography (a qualitative approach), what is important about is not so much that it provides HCI with a method of observation that is attuned to the social world but the "ideas it offers for thinking about social life", it is the "connection at an analytic level" that matters (Dourish, 2006). The same applies to qualitative research more generally. HCI needs to connect at an *analytic level* and appreciate that the ontology, epistemology and methodology of qualitative research – i.e., its world view, its ways of knowing and the logics that provide its methods with their sense and reference – are fundamentally different to and incongruent with deeply ingrained ways of thinking in HCI (Clarke et al. 2025).

There is no S in HCI for science, and there is little if any science in positivism. It is a philosophy of science – a way of thinking about science – not scientific practice, which is a very different beast (Garfinkel, 2022). The idea that HCI should be a science arguably tracks back to Herbert Simon's ambition to create a science of design (Simon, 1996), and leading figures in the field are responsible for promulgating the outmoded idea that science is by definition quantitative in nature, e.g.,

> "Mature scientific domains, such as physics and chemistry, are more likely to have rigorous quantitative laws and formulas, whereas newer disciplines, such as sociology, or psychology, are more likely to have qualitative frameworks and models." Bederson and Schneiderman (2003)

It should be very clear by now that the difference between the quantitative and qualitative has *nothing* to do with new or old, mature or young, nor between natural science and human science. It's entirely to do with what a *human* science should be: a mimic aping a philosophical version of natural science or something that treats the human world and human conduct on its terms. As Dilthey put it, "Nature we explain, psychic life we understand." There is a H for human in HCI and it is foundational. This paper has sought to outline the foundations of human science in order to sensitise the HCI community and reviewers alike to its distinctive character, and to show that aping natural science and trying to make interpretations of the human world accountable to extraneous numerical criteria will not do the job. It's not science, it's scientism, and it suppresses ongoing efforts to understand the relationship of computing technology to the human world through the widespread use of positivist evaluation criteria that put emphasis on quantifiable measures. As previous research has demonstrated (Crabtree et al., 2024), a great deal may be had from qualitative studies that last days, hours, and even minutes. The concept of 'unremarkable computing' (Tolmie et al., 2002) is based on a mere two minutes of data, for example, but it had a significant impact on ubicomp and the notion of 'invisible' computing in HCI. It's not the quantity of data or some other metric that matters, it's the *quality* of insight provided and what it *sensitises* us to that counts. This is not an argument against quantification, surveys and statistics are useful as Douglas (1970) reminds us. It as argument against the inappropriate use of positivistic reasoning. Reviewers may not see themselves as such, indeed they may recoil at the thought, but we are in denial when we dismiss qualitative studies on the basis of how many participants were involved, how long the study was, how much data was gathered, etc., and discount objections to the systemic mistreatment of qualitative research through the imposition of such measures on the basis of the bad luck (that's life) argument. It isn't and it doesn't have to be. The outcome and novel contribution of this crash course in the foundations of human science is *five basic criteria* for evaluating qualitative research in HCI. They furnish reviewers with the tools they need to assure themselves and the HCI community of the quality of interpretive work.

**Acknowledgements.** This work was supported by the Engineering and Physical Sciences Research Council [grant number EP/T022493/1, and EP/V026607/1].